# Approximate Reconstruction of Torsional Potential Energy Surface based on Voronoi Tessellation

Chengming He, Yicheng Chi, and Peng Zhang*

*Department of Mechanical Engineering*

*The Hong Kong Polytechnic University, Hung Hom, Kowloon, Hong Kong*

**Corresponding Author:**
Peng Zhang
Department of Mechanical Engineering
The Hong Kong Polytechnic University
Hung Hom, Kowloon, Hong Kong
E-mail: pengzhang.zhang@polyu.edu.hk
Fax: (852)23654703   Tel: (852)27666664

**Colloquium:**
Gas-phase Reaction Kinetics

**Paper Length (Method 1):**

The total word (excluding title block, abstract and the separate list of figure captions) is: **6185** words

**Word Count** (was performed from automatic counting function in MS Word plus Figures/Equations/References)
**Abstract:**         **183** words, not included in word count

**Main text**:     **3465** words
**Equations:**    **350** words
**References:**   **507** words (27 references)
**Figures:**       **1863** words (8 figures with captions, color figures in electronic version only)

| Figure | Column | Word Count |
|---|---|---|
| 1 | Single | 292 |
| 2 | Single | 172 |
| 3 | Single | 107 |
| 4 | Double | 346 |
| 5 | Single | 248 |
| 6 | Single | 118 |
| 7 | Double | 339 |
| 8 | Single | 241 |
| Total |  | **1863** |

**Total:**   **6185** words

One Supplemental Material is available.

# Approximate Reconstruction of Torsional Potential Energy Surface based on Voronoi Tessellation

Chengming He, Yicheng Chi, and Peng Zhang[*]

*Department of Mechanical Engineering*

*The Hong Kong Polytechnic University, Hung Hom, Kowloon, Hong Kong*

**Abstract:** Torsional modes within a complex molecule containing various functional groups are often strongly coupled so that the harmonic approximation and one-dimensional torsional treatment are inaccurate to evaluate their partition functions. A family of multi-structural approximation methods have been proposed and applied in recent years to deal with the torsional anharmonicity. However, these methods approximate the exact "almost periodic" potential energy as a summation of local periodic functions with symmetric barrier positions and heights. In the present theoretical study, we illustrated that the approximation is inaccurate when torsional modes present non-uniformly distributed local minima. Thereby, we proposed an improved method to reconstruct approximate potential to replace the periodic potential by using information of the local minima and their Voronoi tessellation. First, we established asymmetric barrier heights by introducing two periodicity parameters and assuming that the exact barrier positions are at the boundaries of Voronoi cells. Second, we used multiplicatively weighted Voronoi tessellation to refine the barrier heights and positions by defining a structure-related distance metric. The proposed method has been tested for a few higher-dimensional cases, all of which show promising improved accuracy.

---

[*] Corresponding author
E-mail address: pengzhang.zhang@polyu.edu.hk (P. Zhang)
Fax: (852)23654703    Tel: (852)27666664

# 1 Introduction

Accurate evaluation of partition functions is critical to thermochemical and kinetic calculations of complex molecules[1-4]. This is because torsional modes within a complex molecule containing various functional groups are often strongly coupled (SC), and the resulting torsional anharmonicity[5, 6] renders the harmonic approximation[7, 8] to be highly inaccurate. A variety of studies[9, 10] employ one-dimensional (1D) internal rotation treatment to replace a harmonic oscillator by a specific torsional mode, but it is often difficult to identify SC torsions with specific normal modes[11]. Then, it spawned the non-separable treatments of mixed torsions. The widely discussed Pitzer-Gwinn approximations[12], Feynman path integrals[13], and Monte Carlo phase space integrals[14] have been used but are usually computational expensive for complex molecules.

In recent years, Zheng *et al.*[6, 15] proposed a family of multi-structural (MS) approximation methods, which were believed to satisfactorily deal with the torsional anharmonicity[16-19]. Among the MS methods, the MS-AS method[6] denoting "multi-structural method including all structures" is of particular interest because it does not require any information about conformational barriers or the paths that connect various structures. Furthermore, the exhaustive conformational structure search for all distinguishable structures is extremely complex and expensive for molecules with a large number of torsions (and hence a large number of conformational structures). Consequently, some cost-effective approximation methods, such as the MS-RS[6] method based on including all conformers generated from a reference structure (RS), a dual-level (a low-level and high-level electronic structure) method[20], and an extended two-dimensional (2D) torsion method[21] were also proposed.

The MS-AS method[6] has been used in many studies[5, 17-19, 22]. For example, Zheng *et al.*[5] studied the hydrogen abstraction reactions of iso-butanol by hydroxyl radical, involving 9, 20, 18, 96, and 16 conformational structures for iso-butanol and the R1a−R1d transition states, respectively. They also showed the significance of considering the anharmonicity of high-frequency modes. Zheng *et al.*[22] studied the C–H bond dissociation processes of n-hexane and iso-hexane with containing 23 and 13 conformational structures in the parent molecules and 14 to 45 conformational structures in each of the seven isomeric products. Monge-Palacios *et al.*[17] studied the isomerization reaction of a six carbon atom Criegee intermediate (C6-CI) catalyzed by formic acid and found that the contribution of the reactant C6-CI conformers to the multi-structural partition function is larger than that of the saddle point conformers. Felsmaan *et al.*[19] applied the MS-AS method to predict partition functions of both reactants and transition states in MP (methyl propanoate) + OH/HO$_2$ reaction systems, and they could successfully predict the rate constants compared with literature data. Shang *et al.*[18] used the MS-AS method to investigate the reaction kinetics of H-abstractions from DMA by H, CH$_3$, OH, and HO$_2$ radicals in a broad temperature range (100–2000 K).

We noted that the torsional anharmonicity results in an "almost periodic function" of potential energy on the parameter space of dihedral angles corresponding to the torsional modes. The MS-AS method invokes an essential approximation by expressing such a potential energy function as a summation of several local periodic functions. It means that the constructed potential energy surface (PES) along a specific dihedra angle is always symmetric on two sides of each local minimum. This approximation becomes increasingly inaccurate when torsional modes present non-uniformly distributed local energy minima. In the present study, we attempt to theoretically analyze the uncertainties of the MS-AS method and to propose improved methods for reconstructing approximate PES based on the mathematical technique of Voronoi tessellation.

**2 Theoretical Methodology**

To deal with the torsional anharmonicity induced by SC torsional modes, the MS-AS method[6] uses Voronoi tessellation[6, 23] to identify the influence region for each conformational structure on the parameter space of dihedral angles so as to circumvent the difficulties of assigning coupled torsions to specific normal modes. As shown in Fig. 1, Voronoi tessellation cuts the entire parameter space into several one-by-one subspaces to each local minimum (solid points) and guarantees every point in a subspace is nearest to its corresponding local minimum than other minima. It can be mathematically described as that, for each structure (point) $j$, its influence region $R(j)$ is defined as

$$R(j) = \{x \in X | d^0(x,j) \leq d^0(x,k), \text{for all } j \neq k\} \quad (1)$$

where $x$ is a point in the parameter space $X$, and $d^0$ the distance metric. The Euclidean norm is generally used as the default distance metric. The periodic replicas are also included in the tessellation calculation so that could properly handle the periodic nature of the torsional modes. To physically reflect the influence region for each structure, a multiplicatively weighted Voronoi diagram[23] will be used in the present study. More details about the mathematical theory of Voronoi tessellation can be found in[23, 24].

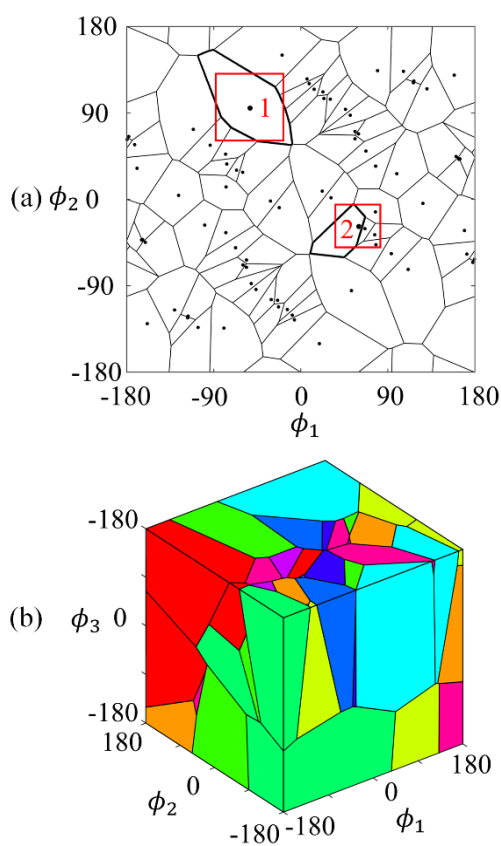

Figure 1. Schematic (a) 2D and (b) 3D Voronoi tessellation generated from the presently calculated 78 conformational structures of MB (Methyl Butanoate) with $HO_2$ in the transition state.

Voronoi tessellation and reconstruction of PES require information from electronic structure calculations. In the present study, all conformational structures with local potential energy minima can be identified by internal torsions from an initial structure (generally a lowest energy state). Density functional theory (DFT) employing the B3LYP functional with the 6-311++G(d,p) basis set[25] was used for geometry optimization, frequency calculation, zero-point energy correction, and hindrance potential treatment. All the calculations were performed with the Gaussian 09 program package.

## 3 Results and Discussions

3.1 Uncertainty analysis of MS-AS method

In the MS-AS method[6], the quantum-mechanical torsional partition function is approximated by a classical mechanical (CM) configuration integral, in which the CM torsional partition function[26] is given by

$$Q^{\mathrm{CM}} = \left(\frac{1}{2\pi\beta\hbar^2}\right)^{t/2} (\det\{\mathbf{D}\})^{1/2} \int_0^{2\pi/\sigma_1} \cdots \int_0^{2\pi/\sigma_t} d\phi_1 \ldots d\phi_t e^{-\beta V(\phi_1,\ldots,\phi_t)} \quad (2)$$

where $\beta = 1/k_\mathrm{B}T$, $k_\mathrm{B}$ the Boltzmann's constant, $T$ the temperature, $\hbar$ the Planck's constant, $\mathbf{D}$ the torsional kinetic energy matrix that evaluated at the global minimum[27], $\phi_\tau$ the torsional internal coordinate, $\sigma_\tau$ the parameter characterizing the periodicity of torsional space, and $t$ the total number of coupled torsions.

Owing to the existence of a large number of local potential energy minima that contribute to the evaluation of partition function, the entire torsional space can be topographically divided into a number of distinct subspaces corresponding to each local minimum (structure), and thereby the total partition function is a summation of that for all the minima (structures). For each structure $j$ belonging to a certain torsion $\tau$, the PES is assumed to be a periodic function:

$$V_{j,\tau} = U_j + \frac{W_{j,\tau}}{2}[1 - \cos M_{j,\tau}(\phi_\tau - \phi_{\tau.\mathrm{eq}.j})], \qquad \frac{-\pi}{M_{j,\tau}} \leq \phi_\tau - \phi_{\tau.\mathrm{eq}.j} \leq \frac{\pi}{M_{j,\tau}} \quad (3)$$

where $U_j$ and $\phi_{\tau,\mathrm{eq},j}$ are respectively the potential energy and the torsional internal coordinate of the local minimum $j$, $M_{j,\tau}$ the periodicity parameter. $W_{j,\tau}$ is the effective barrier height estimated from $M_{j,\tau}$, the frequency $\omega_{j,\tau}$, and the internal moment of inertia $I_{j,\tau}$ by

$$W_{j,\tau} = \frac{2I_{j,\tau}\omega_{j,\tau}^2}{M_{j,\tau}^2} \tag{4}$$

Furthermore, for the torsional anharmonicity induced by SC torsions, the PES within a certain structure $j$ is assumed to be separable so that

$$V_j(\phi_1, \dots, \phi_t) \approx \sum_{\tau=1}^{t} V_{j,\tau}(\phi_\tau) \tag{5}$$

where each $V_{j,\tau}(\phi_\tau)$ is calculated by Eq. (3).

The assignment of $M_{j,\tau}$ is a crucial part of the MS-AS method. For nearly separable (NS) torsions with approximately evenly distributed local minima, $M_{j,\tau}$ simply equals to the total number of local minima in the specific torsion; whereas for SC torsions, as shown in Fig. 1, $M_{j,\tau}$ is determined by Voronoi tessellation[6], in which $M_{j,\tau}$ is replaced by $M_j^{SC}$ and assumed to be equivalent in every SC torsion. Then, the local periodicity $M_{j,\tau}$ is defined as

$$M_{j,\tau} = M_j^{SC} = \frac{2\pi}{\left(\Omega_j^{SC}\right)^{1/t_{SC}}} \tag{6}$$

where $\Omega_j^{SC}$ is the hypervolume of subspace $j$ and $t_{SC}$ is the total number of SC torsions. Overall, $M_{j,\tau}$ plays three important roles in the method. First, it controls the local periodicity. Second, it determines the integral subspace for a specific structure. Third, it accounts for the evaluation of implicit barrier heights.

The periodic potential assumption of Eq. (3) in the MS-AS method is conditionally accurate if the local minima are uniformly distributed with same energies and frequencies. As the 1D example shown in Fig. 2, the potential $V_1/\text{cm}^{-1} = 750\cos(2\phi) + 750$ is a periodic function with the periodicity of 180 degree. Then it can be divided into two identical periodic functions that are corresponding to two local minima, respectively, in which each periodic function traverses a complete period of 180 degree.

The exact potential is generally not a periodic function if existing the torsional anharmonicity, as the 1D torsional potential[6] for $H_2O_2$ shown in Fig. 2, that the potential $V_2/\text{cm}^{-1} = 830.7 +$ $1037.4\cos(\phi) + 674.2\cos(2\phi) + 46.9\cos(3\phi) + 2.7\cos(4\phi)$ is an almost periodic function with a permanent periodicity of 360 degree because of the natural torsion. The MS-AS method splits the potential $V_2$ into two piecewise periodic functions $V_3$ that corresponding to two local minima. According to Eq. (3) and (4) with the periodicity parameter $M = 2$, the internal moment of inertia $I = 0.4232$ amu Å$^2$, and the harmonic frequency $\omega = 382.6$ cm$^{-1}$ for both two minima, the reconstructed potential $V_3$ is also plotted in Fig. 2. It is seen that the potential curve $V_3$ is always symmetric on two sides of the local minimum with same barrier heights, which causes the potential to be one side higher and the other side lower than the exact potential. It is also found an overlapped region between two neighboring periodic functions. Consequently, the periodic potential assumption properly leads to inaccurate estimation of the partition function, which is believed to be increasingly exacerbated when the local minima become more non-uniformly distributed.

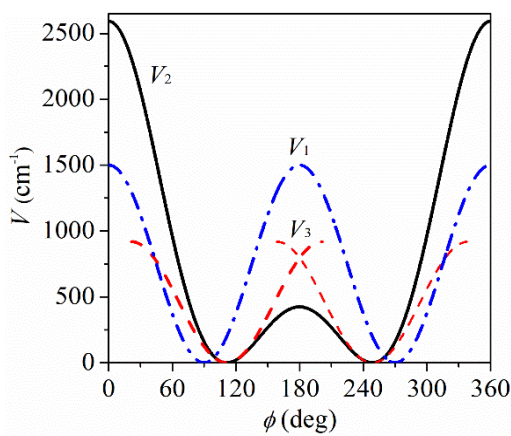

Figure 2. Potentials of a periodic function $V_1$, an almost periodic function $V_2$ (corresponding to H$_2$O$_2$), and piecewise periodic functions $V_3$ that constructed in the MS-AS method.

The uncertainty of the MS-AS method becomes prominent when considering SC torsions. As the red frame of representative structures 1 and 2 shown in Fig. 1(a), the constructed subspace in the MS-AS method is a square for 2D and a cube for 3D coupled torsions. From a closeup of structure 1 shown in Fig. 3, we can infer that the Voronoi subspace and the constructed MS-AS subspace contribute approximately equally on the partition function for those structures with local minimum located almost centrally. However, for those structures whose local minima are very close to the boundary of their Voronoi cells, for example structure 2, the periodic potential assumption in the MS-AS method would induce highly inaccurate partition function estimation. This can be understood by a closeup of structure 2 shown in Fig. 3. Apart from the overlapped white region, the shade region denoted by vertical lines of isotropic subspace is closer to the local minimum than that denoted by horizontal lines of Voronoi subspace, and thereby has a lower potential energy $V$ and a high partition function $Q \sim e^{-V/k_\text{B}T}$. It is predicted that the deviations of partition function estimation between the MS-AS method and the exact value would be increasingly enlarged when considering three SC torsions in the transition state involving extremely non-uniformly distributed local minima. For example, Li *et al.* [16] found that the multi-structural torsional partition functions are 5.67 to 11.81 times larger than the single-structural partition functions for the transition state of the hydrogen abstraction reaction from methyl butenoate by H atoms.

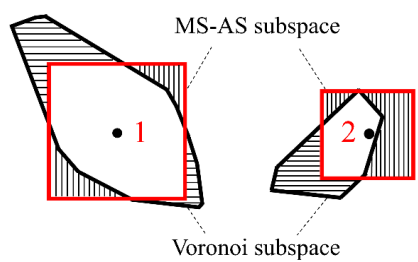

Figure 3. Closeup of Voronoi subspace and constructed MS-AS subspace in Fig. 1.

3.2 Improved reconstruction of torsional PES

Considering the uncertainties of the MS-AS method, we considered to relax the assumption of the periodic potential and to obtain a physically reliable potential. We noted that a higher-level MS method (MS-ASCB)[6] has been introduced and ASCB denotes "based on all structures and conformational barriers". It includes the explicit conformational barrier heights and positions obtained from electronic structure calculations, and hence the continuous torsional potential can be given as

$$V_{j,\tau} = \begin{cases} U_j + \dfrac{W_{j,\tau}^L}{2}\left[1-\cos\left(\dfrac{(\phi_\tau - \phi_{\tau.\text{eq}.j})\pi}{(\phi_{\tau.\text{eq}.j} - \phi_{j,\tau}^L)}\right)\right], (\phi_{j,\tau}^L \leq \phi_\tau \leq \phi_{\tau.\text{eq}.j}) \\ U_j + \dfrac{W_{j,\tau}^R}{2}\left[1-\cos\left(\dfrac{(\phi_\tau - \phi_{\tau.\text{eq}.j})\pi}{(\phi_{j,\tau}^R - \phi_{\tau.\text{eq}.j})}\right)\right], (\phi_{\tau.\text{eq}.j} \leq \phi_\tau \leq \phi_{j,\tau}^R) \end{cases} \quad (7)$$

where $W_{j,\tau}^L$ and $W_{j,\tau}^R$ are the exact left and right barrier heights along torsion $\tau$ of structure $j$, and $\phi_{j,\tau}^L$ and $\phi_{j,\tau}^R$ are their corresponding barrier positions. Theoretically, the MS-ASCB method can provide more reliable potential than the MS-AS method. But, the additionally required information of barrier heights and positions would cause a significant amount of computational cost and human efforts. Consequently, it is quite compelling to think of a more physical potential to replace the periodic potential in the MS-AS method because the information of barrier positions and heights are reflected by the shape and boundary of the Voronoi tessellation.

We attempt to reconstruct an almost periodic function of potential based on the Voronoi tessellation in two steps. First, we assume that the barrier positions are at the boundaries of Voronoi cells, and then calculate the barrier heights according to Eq. (4) by defining another two periodicity parameters, $M_{j,\tau}^L$ and $M_{j,\tau}^R$, which are related to $M_{j,\tau}$. As a result, we can characterize the asymmetric barrier positions and heights for the local minimum. Second, recognizing that barrier positions tend to vary with the energy and frequency of the local minimum, we define a structure-related distance metric to recalculate the Voronoi tessellation and to obtain the corrected barrier positions and heights.

Specifically, the left and right boundaries of Voronoi cells in each torsion, $\phi_{j,\tau}^L$ and $\phi_{j,\tau}^R$, are assumed as the exact barrier positions, which are already determined from the Voronoi tessellation, as the hollow intersection points of blue Voronoi subspace and red MS-AS subspace shown in Fig. 4(a). The local periodicity parameter $M_{j,\tau}$ for each structure is equivalent in every torsion and determined by Eq. (6). Then, to characterize the barrier height difference on the left and right sides of the local minimum in each torsion, we define two periodicity parameters, $M_{j,\tau}^L$ and $M_{j,\tau}^R$, by

$$M_{j,\tau}^L \Omega_{j,\tau}^L = M_{j,\tau}^R \Omega_{j,\tau}^R \quad (8)$$

$$\left(M_{j,\tau}^L\right)^2 + \left(M_{j,\tau}^R\right)^2 = 2M_{j,\tau}^2 \quad (9)$$

where $\Omega_{j,\tau}^L$ and $\Omega_{j,\tau}^R$ are the hypervolumes obtained by cutting the subspace into "left" and "right" parts with a specific plane, in which the specific plane is perpendicular to the torsion axis and across the internal coordinate $\phi_{j,\tau}$ of the local minimum, as shown in Fig. 4(a). Subsequently, we can obtain the corrected barrier heights, $W_{j,\tau}^L$ and $W_{j,\tau}^R$, by Eq. (4), and the almost periodic potential by Eq. (7).

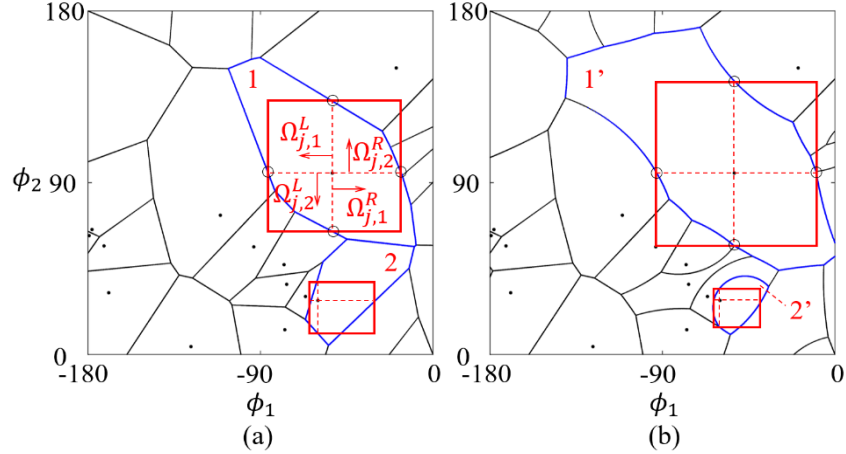

Figure 4. A closeup of upper left corner of Fig. 1(a) showing the improvements of the MS-AS method with corrected (a) barrier heights based on original Voronoi diagram and (b) barrier positions based on multiplicatively weighted Voronoi diagram.

In general cases, the barrier positions are not exactly at the boundaries of Voronoi cells, since the Voronoi tessellation topologically divides the space merely in geometry by assuming all structures contribute equally to the total potential. But in fact, the barrier positions would change as varying the energy and the frequency of each local minimum. Physically, for the ideal states of evenly distributed local minima with same energies and frequencies, the exact barrier positions should be completely located at the boundaries of Voronoi cells. However, for the general states involving frequency and energy differences between neighboring structures, the barrier positions tend to be closer to those minima with higher frequencies or higher energies. Based on this understanding, we define a new structure-related distance metric as $d_{j,\tau}^1 = \varepsilon_{j,\tau} d_{j,\tau}^0$, where $d_{j,\tau}^0$ is the distance metric that used in the original Voronoi tessellation. The correction coefficient $\varepsilon_{j,\tau}$ is defined as

$$\varepsilon_{j,\tau} = \frac{d_{j,\tau}^1}{d_\tau^0} = \frac{U_j + \frac{2I_{j,\tau}\omega_{j,\tau}^2}{M_{j,\tau}^2}}{\frac{1}{N}\sum_{j=1}^{N}\left(U_j + \frac{2I_{j,\tau}\omega_{j,\tau}^2}{M_{j,\tau}^2}\right)} \qquad (10)$$

to reflect the synergetic effects of frequency and energy changing on the variation of barrier positions. A weighted correction coefficient $\varepsilon_j$ is defined as the root mean square of all concerned torsions $\varepsilon_{j,\tau}$. Equation (10) shows that either increasing $U_j$ or $\omega_{j,\tau}$ would result in an increased $\varepsilon_{j,\tau}$ and $d_{j,\tau}^1$, and thereby a reduced hypervolume of Voronoi subspace $j$. This can be clearly shown by the multiplicatively weighted Voronoi diagram in Fig. 4(b) that the Voronoi cells for structure 1' ($\varepsilon_1 = 0.5$) and 2' ($\varepsilon_2 = 1.3$) are enlarged and reduced, respectively. Furthermore, for any two neighboring local minima with same $U_j$ and $\omega_{j,\tau}$, their $\varepsilon_{j,\tau}$ are the same (but not has to be 1) so that the original boundaries of Voronoi cells are not changed.

We name the improved method as MS-ASB, where B denotes the corrected "barrier heights and positions" by information of Voronoi tessellation. It is noted that, for 1D case, the potential integration of Eq. (2) is performed in Voronoi subspace rather than the constructed isotropic subspace in the original MS-AS method. Whereas, for high dimensional cases, the potential integration is performed in a rectangle subspace, as the red frame shown in Fig. 4(a), rather than the Voronoi subspaces (arbitrary polygon) owing to the separability assumption of potential in each coupled torsion. To facilitate the application of the proposed MS-ASB method, the 1D example in a specific torsion is used to clearly present the derivation of asymmetric barrier heights and weighted Voronoi subspace in the Sec. 1 of the Supplemental Material, and then the high-dimensional coupled torsions can be thereby obtained in a straightforward manner.

3.3 Testing cases

In the present study, the improved MS-ASB method was tested by a number of cases, such as 1D torsional potentials of $H_2O_2$ and a toy model, ethanol radical, MB+$HO_2$, 1-pentyl radical, and 1-butanol radical. The process of partition function calculations and computational data used in the test cases have been included in the Sec. 2 and 3 of the Supplemental Material, respectively. The MATLAB code for these test cases are also attached so that interested readers can use it to reproduce all the results. Here, a 1D torsional potential of $H_2O_2$, a 2D torsional mode of 1-pentyl radical, and a 3D torsional mode of 1-butanol radical were chosen for better illustration.

In the first testing case, the 1D torsion potential[6] of $H_2O_2$ that has been discussed in Fig. 2 is used. Owing to the symmetric distribution of local minima, the exact barrier positions are just at the boundaries of Voronoi subspaces. As shown in Fig. 5(a), the reconstructed almost periodic function of potential by the MS-ASB method is more physically reliable than the symmetric potential by the MS-AS method. Figure 5(b) shows the partition function ratio (PFR) normalized by the exact value of partition function. The MS-AS method overestimates the partition function. However, the improved MS-ASB method can reduce the overestimation by a factor of 1.2 and obtain an accurate estimation of partition function that closer to the exact value particularly at high temperatures, regardless of some deviations at low temperature. This might be attributed to the slightly larger barrier height estimated by MS-ASB.

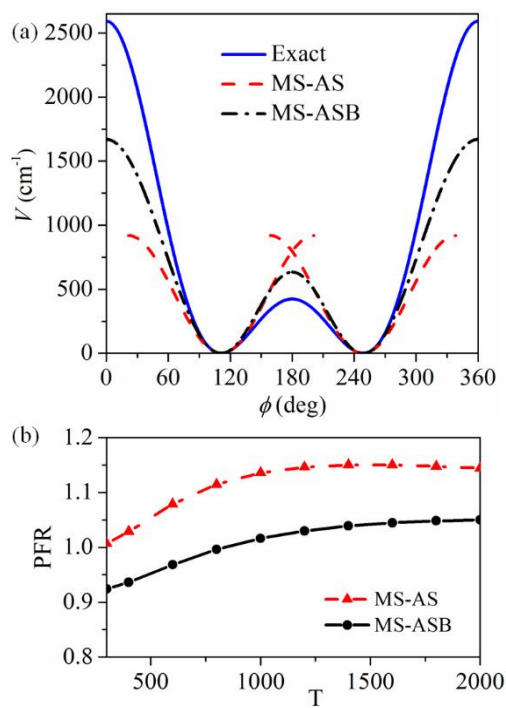

Figure 5. Comparison of (a) the potential curve and (b) partition function ratio (PFR) of the 1D torsion

potential[6] of $H_2O_2$.

In the second testing case, the 2D torsional mode of 1-pentyl radical[6] with 15 conformational structures is used. Figure 6 shows the comparison of partition function $Q^{MS\text{-}ASB}/Q^{MS\text{-}AS}$ due to the lack of exact value. The calculation results by using the MS-ASCB method is however provided neither in[6] nor subsequent studies[15, 20, 21]. The ratio $Q^{MS\text{-}ASB}/Q^{MS\text{-}AS}$ is smaller than the unity, which is consistent with the theoretical analysis that the MS-AS method tend to predict a larger value of PF if there are many non-uniformly distributed local minima. This is because, geometrically, the weighted Voronoi tessellation in the MS-ASB method can modify the original "banded" Voronoi subspace in the MS-AS method to a more "round" subspace with the local minima located centrally, as shown in Fig. 7.

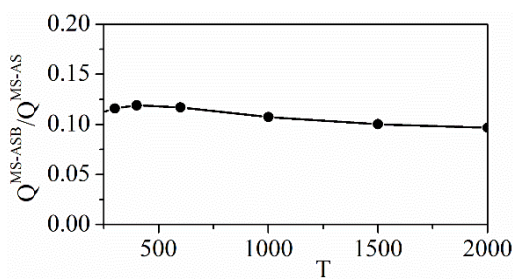

Figure 6. Comparison of partition function $Q^{MS\text{-}ASB}/Q^{MS\text{-}AS}$ of the 2D torsional mode of 1-pentyl radical.

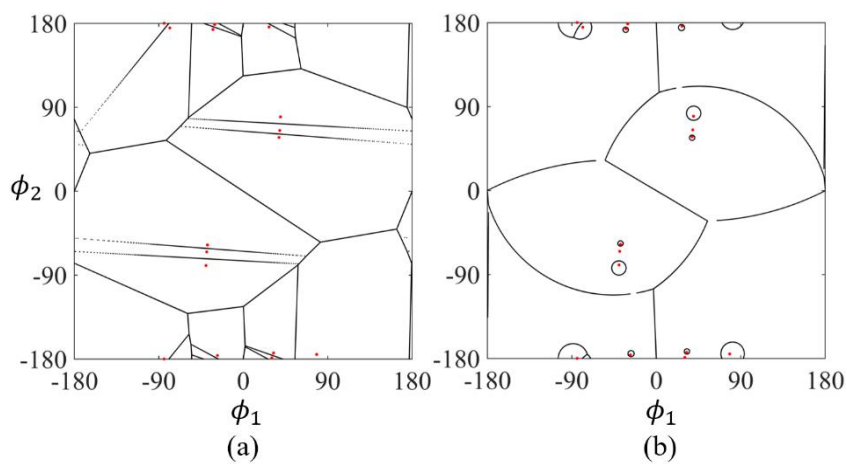

Figure 7. Schematic of the (a) original Voronoi diagram in the MS-AS method and the (b) weighted

Voronoi diagram in the MS-ASB method for the 2D torsional mode of 1-pentyl radical.

In the third testing case, the 3D torsional mode of 1-butanol radical[6] with 29 conformational structures is used. It is seen that, as show in Fig. 8(a), the ratio $Q^{MS\text{-}ASB}/Q^{MS\text{-}AS}$ is close to the unity although there are more coupled torsions than that in Fig. 6. This is probably owing to the relatively uniform distribution of the local energy minima shown in Fig. 8(b). It implies that the improvement of the MS-ASB method mainly relies on the largely non-uniform distribution of the local minima as discussed in Fig. 7 rather than by considering more coupled torsions.

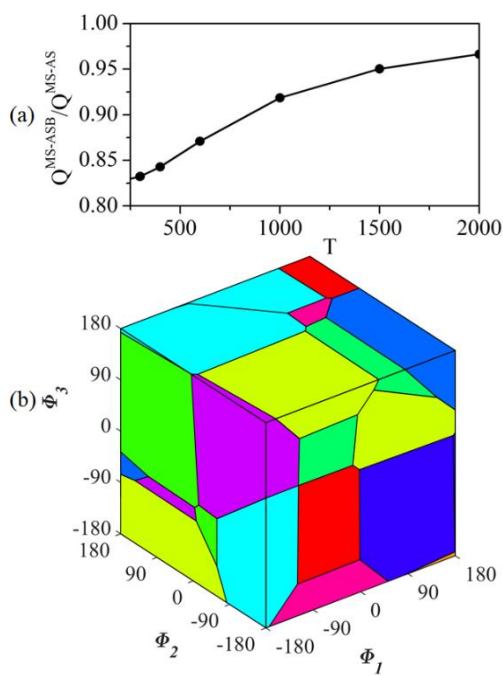

Figure 8. Test of the 3D torsional mode of 1-butanol radical. (a) comparison of partition function $Q^{MS\text{-}ASB}/Q^{MS\text{-}AS}$ and (b) schematic of weighted Voronoi diagram.

In summary, both the theoretical analysis and numerical results show that the MS-AS method tends to predict a larger value of PF, but the proposed MS-ASB method can reduce the possible overestimation by a factor that depends on the specific chemical systems. It is also worthy to mention that, for the 2D torsional mode of ethanol radical in the Supplemental Material, the ratio $Q^{MS\text{-}ASB}/Q^{MS\text{-}AS}$ is nearly the unity because the conformational structures are highly symmetric, which indicates that the MS-ASB method can be degenerated to the MS-AS method for the evenly distributed local minima so that to further consolidate the validations.

Furthermore, the proposed MS-ASB method does not cause much additional cost because all the parameters used in it are obtained from the Voronoi Tessellation, which is the required procedure for other MS-based methods. These test cases in the present study show that the computational cost of the MS-ASB method is about twice of the MS-AS method by using our MATLAB code, which has been provided in the Supplemental Material. The most time-consuming process involved in the MS-ASB method is the calculation of $M_{j,\tau}^L$ and $M_{j,\tau}^R$ by the Monte Carlo sampling method.

## 4 Conclusions

Motivated by the observations that the widely-used MS-AS method to deal with the torsional anharmonicity often overpredicts partition functions of complex molecular system, we performed an uncertainty analysis to show that its periodic potential assumption would result in inaccurate estimation of partition function when torsional modes present non-uniformly distributed local energy minima. To remedy this problem, we proposed the improved MS-ASB method to reconstruct an almost periodic function of potential to replace the periodic potential based on the Voronoi tessellation of local minima. The testing cases in the present study show that the MS-ASB method is promising to predict more accurate PF compared with the MS-AS method when addressing the torsional anharmonicity problem with extremely non-uniformly distributed local minima. The proposed MS-ASB method can be treated as an intermediate method, which aims to have the comparable computational cost to the MS-AS method but to have computational accuracy approximate to that of the MS-ASCB method. Separate future works are merited to further validate the MS-ASB method against the MS-ASCB method and other higher-level theoretical methods, which often require the very large amount of computational costs for exact barrier heights and positions of couple torsional modes.

**Acknowledgment**

This work was supported by NSFC (91641105) and partly by the university matching grant (4-BCE8) and DGRF (G-UAHP).

**List of Figure captions**

(Color figures in electronic version only)

Figure 1. Schematic (a) 2D and (b) 3D Voronoi tessellation generated from the presently calculated 78 conformational structures of MB (Methyl Butanoate) with $HO_2$ in the transition state.

Figure 2. Potentials of a periodic function $V_1$, an almost periodic function $V_2$ (corresponding to $H_2O_2$), and piecewise periodic functions $V_3$ that constructed in the MS-AS method.

Figure 3. Closeup of Voronoi subspace and constructed MS-AS subspace in Fig. 1.

Figure 4. A closeup of upper left corner of Fig. 1(a) showing the improvements of the MS-AS method with corrected (a) barrier heights based on original Voronoi diagram and (b) barrier positions based on multiplicatively weighted Voronoi diagram.

Figure 5. Comparison of (a) the potential curve and (b) partition function ratio (PFR) of the 1D torsion potential[6] of $H_2O_2$.

Figure 6. Comparison of partition function $Q^{MS-ASB}/Q^{MS-AS}$ of the 2D torsional mode of 1-pentyl radical.

Figure 7. Schematic of the (a) original Voronoi diagram in the MS-AS method and the (b) weighted Voronoi diagram in the MS-ASB method for the 2D torsional mode of 1-pentyl radical.

Figure 8. Test of the 3D torsional mode of 1-butanol radical. (a) comparison of partition function $Q^{MS-ASB}/Q^{MS-AS}$ and (b) schematic of weighted Voronoi diagram.

**Supplemental Material**

| Item | File name | Content (captions) |
|---|---|---|
| 1 | PROCI-D-19-01430R2-Supplemental Materials.docx | 1. Derivation of asymmetric barrier heights and weighted Voronoi subspace 錯誤! 尚未定義書籤。<br>2. Process of partition function calculations by attached MATLAB codes .. 錯誤! 尚未定義書籤。<br>2.1. Input files for the *Run_1_preperation.m* executable 錯誤! 尚未定義書籤。<br>2.2. Input files for the *Run_2_weighted_voronoi.m* executable 錯誤! 尚未定義書籤。<br>2.3. Input files for the *Run_3_pf.m* executable ............. 錯誤! 尚未定義書籤。<br>2.4. Input files for the *2D_PES.m* executable ............... 錯誤! 尚未定義書籤。<br>3. Computational data for testing cases ........................ 錯誤! 尚未定義書籤。<br>3.1. Test case of 1D torsion potential $H_2O_2$ ................. 錯誤! 尚未定義書籤。<br>3.2. Test case of 1D artificial torsional model .................... 錯誤! 尚未定義書籤。<br>3.3. Test case of 2D torsional mode of ethanol radical 錯誤! 尚未定義書籤。<br>3.4. Test case of 2D torsional mode of MB+$HO_2$ radical 錯誤! 尚未定義書籤。<br>3.5. Test cases of 2D and 3D torsional modes of 1-pentyl radical 錯誤! 尚未定義書籤。<br>3.6. Test case of 3D torsional mode of 1-butanol radical 錯誤! 尚未定義書籤。<br>4. Summary .................................................................. 錯誤! 尚未定義書籤。<br>Reference ...................................................................... 錯誤! 尚未定義書籤。 |

| 2 | PROCI-D-19-01430R2-matlab_codes.zip | It contains 7 folders: /matlab_codes/H2O2 /matlab_codes/artificial_model /matlab_codes/ethanol /matlab_codes/MB+HO2 /matlab_codes/1-pentyl-2d /matlab_codes/1-pentyl-3d /matlab_codes/1-butanol-3d  Each folder has four executables: *Run_1_preperation.m* *Run_2_weighted_voronoi.m* *Run_3_pf.m* *2D_PES.m* |
|---|---|---|